\documentclass[referee]{raa}            

\usepackage{graphicx,times}             

\begin{document}

\title{ The mid-infrared variability of the SDSS optical quasars  }

\volnopage{ {\bf 2019} Vol.\ {\bf X} No. {\bf XX}, 000--000}
   \setcounter{page}{1}

\author{Hongtao Wang\inst{1,2,3}, Yong Shi\inst{1,3}$^{\dag}$  }

   \institute{School of Astronomy and Space Science, Nanjing University, Nanjing 210093, China:   
   {\it yong@nju.edu.cn} \\
     \and
           The School of Science, Langfang Normal University, Langfang 065000, China
            \\    
     \and
     Key Laboratory of Modern Astronomy and Astrophysics, Nanjing University, Nanjing
210023, China   \\                 
{\small Received ********; accepted ********}}

\abstract{ Based on the Seventh Data Release (DR7) quasar catalog from the Sloan Digital Sky Survey, we investigate the variability of optical quasars in W1, W2, W3 and W4 bands of the Wide-field Infrared Survey Explorer (WISE) and the Near-Earth Object Wide-field Infrared Survey Explorer (NEOWISE). Adopting the structure function method, we calculate the structure function ($\rm\delta t$=1 yr) which shows no obvious correlations with the bolometric luminosity, the black hole mass and the Eddington ratio. The ensemble structure functions in W1 and W2 bands show that the SF slopes are steeper than those in previous studies which may be caused by different cadence and observational epoch number. We further investigate the relation of variability amplitude $\sigma_m$ between mid-infrared band and optical band, but no obvious correlation is found. No correlation is found between W1-W2 and g-r color. We think the mid-infrared emission of quasars may be smoothed out by the extended dust distribution, thus leading to no obvious correlation. For the radio-loud quasar sub-sample, we further analyze the relation between the  variability amplitude in the mid-infrared band and the radio luminosity at 6 cm, 
but no obvious correlations are found, which indicate the mid-infrared emission contributed from the synchrotron radiation of the relativistic jet is very weak.   
}

   \authorrunning{Hongtao Wang \& Yong Shi }            
   \titlerunning{  The mid-infrared variability of the SDSS optical quasars      }  
   \maketitle

\section{Introduction}
 Active galactic nuclei (AGNs) are powered by the supermassive black holes (SMBHs) that accrete matter and release enormous energy (Rees 1984). Surrounding the central SMBH, there is usually an accretion disc and a dusty torus (Urry \& Padovani 1995), which account for the observed big blue bump (BBB) and mid-infrared bump in their spectral energy distribution (SED), respectively. The BBB is considered as the thermal emission from the accretion disc (Shakura \& Sunyaev 1973) and the mid-infrared bump is believed to originate from the dusty torus (Koshida et al. 2014; Mandal et al. 2018). 

Quasars have long been recognized as intrinsically variable sources since their discovery ((Matthews \& Sandage 1963; Wagner \& Witzel 1995; Ulrich, Maraschi \& Urry 1997). The aperiodic flux variability of AGNs is ubiquitous from radio to high-energy $\gamma$-ray, and the timescales vary from minutes to years (Ulrich et al. 1997). The variability provides us many invaluable clues to study AGNs, such as estimating the black hole mass (Peterson et al. 2004; Vestergaard \& Peterson 2006 and reference therein), constraining the structure of broad line regions (Pancoast et al. 2011) and identifying AGNs (Padovani et al. 2017) etc.   

In mid-infrared (mid-IR) band, the emission is dominated by the dusty torus (see, e.g., Nenkova et al. 2008), which reprocesses the UV/optical emission from accretion disks into the mid-IR emission. Non-thermal emission from jets sometimes also contributes to some fractions of the mid-IR luminosity in radio-loud AGNs. Based on 4$\sim$5 epoch observations, Koz{\l}owski et al. (2016) analyzed the ensemble mid-IR variability of $\sim$ 1500 quasars by the single power-law structure function (SF), and found the SF slope in the mid-IR band was steeper than that in the optical band. Moreover, the SF slope show no correlations with the luminosity in mid-IR band and rest-frame wavelength, while the amplitude does show a correlation with the rest-frame wavelength and an anti-correlation with the mid-IR luminosity. 
 
With the coming of WISE, there is a new opportunity to carry out studies of mid-IR variability. The WISE mission is divided into two main cycles. The first stage is the full solid hydrogen cryogenic phase with W1, W2, W3 and W4 channels and the asteroid-hunting phase with only W1 and W2 channels from January 2010 to February 2011. All observed data sets during this stage are included in the `AllWISE MultiEpoch Photometry (MEP) Database'. The second stage is near earth object WISE Reactivation mission during 2014 to 2018 with W1 and W2 bands. The data sets in this stage are included in the `NEOWISE-R Single Exposure (L1b) Source Table'. The total combined baseline is $\sim$ 9 years in W1 and W2 bands, which offers a chance to investigate the variability characteristics of quasars in mid-infrared band. Comparing with Spitzer studies (Koz{\l}owski, et al. 2016 and reference therein), WISE/NEOWISE offers much higher cadence and larger sample size which provide a great opportunity for us to systematically investigate the mid-infrared variability of quasars.

The paper is organized as follows. The data sets and sample selections are described in Section 2. 
The analytical method and main results are presented in Section 3. The discussion and conclusion of this work  are given in Section 4 and Section 5, respectively. Throughout this paper, we assume a flat $\Lambda$CDM cosmology with $\rm\Omega_M=0.30$, $\rm\Omega_\lambda=0.70$ and $\rm H_0=70.0~km~s^{-1} Mpc^{-1}$.

\section{Sample and Data }   
 In order to investigate the variability of quasars in mid-infrared band, we utilize the DR7 quasar sample from the Sloan Digital Sky Survey (Shen et al. 2011), including 105,783 quasars brighter than 
$\rm M_i= -22.0$ mag. This sample has at least one broad emission line with the FWHM larger than 
$\rm 1000~ km~s^{-1}$, and also contains many available parameters, e.g., redshift, black hole mass, Eddington ratio and  luminosity in 5100 \AA~etc. We cross-match the SDSS DR7 quasars catalogue with the mid-IR catalogue, `AllWISE MultiEpoch Photometry  Database' and `NEOWISE-R Single Exposure (L1b) Source Database'(update to April 2019) using a search radius of 5 arcsec. All the 105,783 SDSS quasars have WISE counterparts. The angular resolution is 6.1\arcsec, 6.4\arcsec, 6.5\arcsec and 12.0\arcsec in W1,W2,W3 and W4 bands, respectively. Our search radius of 5\arcsec is thus in general larger than the position accuracy of WISE sources (angular resolution divided by the signal-to-noise). We exclude the epoch photometry without errors and merge observations within one  day.  

Some bad photometric measurements are further excluded in W1,W2,W3 and W4 bands for these objects with the  criterias as follows. 

\begin{enumerate}    
\item The  number of point spread function (PSF) components in the profile 
fit for a object (\texttt{nb}) must be less than 3, and the  best quality single-exposure image (
qi\underline{\hspace{0.5em}}fact=1) are not actively deblended (na$=$0).
\item The signal noise ratio (SNR) is larger than 10. The reduced $\chi^2$ of single-exposure profile-fit is less than $5$ in both W1 and W2 bands (i.e., w1rchi2$<5$ and w2rchi2$<5$). 
\item Both the contamination and confusion flag (cc\underline{\hspace{0.5em}}flags) are labeled as 0000. The  angular distance from the nominal boundaries of the South Atlantic Anomaly (saa\underline{\hspace{0.5em}}sep) is larger than 5.0. Moon masking flag (moon\underline{\hspace{0.5em}}masked) is setted as 00 to avoid contamination by scattered moonlight. Overall frameset quality score (qual\underline{\hspace{0.5em}}frame) is larger than 0 to ensure the framesets with good quality.  
\end{enumerate}
  
We also apply a 5$\sigma$ criteria ($\sigma$ is the standard deviation of the light curve) to clip each light curve for removing the photometric outliers which may be caused by cosmic-rays. Finally, 103,762 objects at least two epochs in W1 band and W2 band are remain to investigate the long-term variability. The distribution of 
 bolometric luminosity and redshift of the final sample is shown in Figure 1. 
In our final sample, the minimum, median and maximum time span of the observation is $\sim$ 0.5 years, 4.5 years and 8.5 years, respectively. The minimum, median and  maximum epoch number are 2, 6 and 143, respectively. We also utilize the same criteria to construct the light curve in W3 and W4 bands but with much shorter baseline which is 9 months and 7 months, respectively.

\begin{figure}[htbp]
\centering  
\resizebox{8.3cm}{6.3cm}{\includegraphics{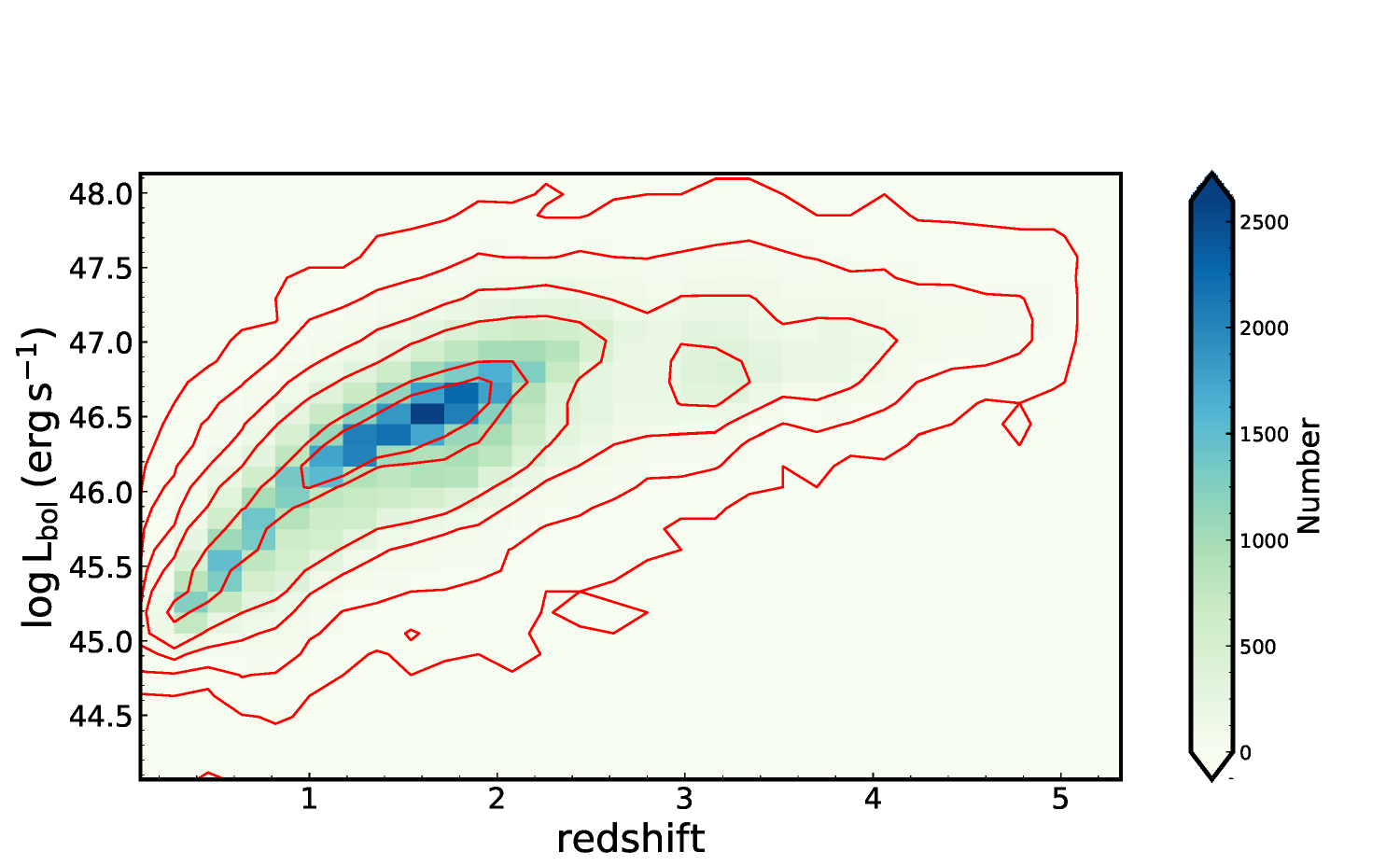}}
\caption{ The distribution of bolometric luminosity and redshift of the final 
quasar sample.}\label{Fig:radio_var}. 
\end{figure} 
   
\section{ The methods to characterize the variability  }
\subsection{ Structure Function Method }            
The structure function (SF) method is a typical technique to analyze the variability of AGNs. It calculates the  rms amplitude as a function of time lag $\delta t$ between different observations. SF is not affected by   windows problem and aliasing due to sparse sampling (Hughes et al. 1992), and suitable for irregular data sets. Among different definitions of SFs (Simonetti, et al. 1985; Vanden Berk et al. 2004; Sumi et al. 2005; Bauer et al. 2009; MacLeod et al. 2012), we utilize the one by Koz{\l}owski et al. (2016)  
\begin{equation} 
SF_{\rm obs}(\delta t)  =  \sqrt{\frac{1}{N}\sum_{i=1}^{N} (y(t)-y(t+\delta t))^2},  \\	  
\label{eq:sfrms}
\end{equation}
\noindent  
where $y(t)$ is the observed flux at time t, N is the total observation pairs in the time lag bin $\delta t$. 
SF is calculated by three processes as follows. First, we calculate the time lag and magnitude variation of each pairs in the data sets, and sort the results by time lag. 
Second, we correct the time lag in observed frame $|t_i - t_j|$ into rest frame by $\delta t = |t_i - t_j|(1+z)^{-1}$. Third, we divide the $\delta t$ into several bins and calculate the SF in each bin by expression (1). The bin size is depend on the number in the data sets. The target is to obtain the well-distributed SF as far as possible. Emmanoulopoulos et al.(2010) pointed out that the limited time sampling of individual object could give rise to spurious features in the SF result, but it could be relieved by the ensemble SF (Sun, et al. 2015).           
The SF in each bin is calculated by the bootstrap method 1000 resampling and choose the $1\sigma$ value as the error of SF.

\subsection{ The variability amplitude } 
Besides the SF, the variability amplitude ($\sigma_m$) is also usually used to investigate the variability of quasars. The amplitude measures the variance of the light curve after correcting the measurement uncertainty (Sesar et al. 2007; Ai et al.2010; Ai et al.2013; Rakshit \& Stalin 2017; Rakshit et al. 2019). The $\sigma_m$ is calculated as the following processes. The standard deviation ($\Sigma$) of the observed signal in the data sets is first given by    
\begin{equation}
\Sigma=\sqrt{\frac{1}{n-1}\sum_{i=1}^{n}(m_i - <m>)^2},
\end{equation}
in which $m_i$ is the magnitude of $i$-th epoch, and $<m>$ is the mean amplitude, n is the number of detections. The error $\epsilon$ is then calculated from the individual error with the following expression 
\begin{equation}
\epsilon^2=\frac{1}{N}\sum_{i=i}^{N}{\epsilon_{i}^{2} + \epsilon_{s}^2}. 
\end{equation}
Here $\epsilon_{i}$ is the uncertainty of $i$-th measurement value and $\epsilon_{s}$ is the systematic uncertainty. Jarrett et al. (2011) showed the systematic uncertainties of 0.024 mag and 0.028 mag in $\rm W$1 and $\rm W$2, respectively. At last, the variability amplitude can be written as  
            
\begin{equation}
\sigma_m  = {\sqrt{\Sigma^2 - \epsilon^2}  ,   \quad    \Sigma>\epsilon } , 
\end{equation}
otherwise, the $\sigma_m =0$. 

\section{ The results }  
\subsection{ The ensemble structure function in W1,W2,W3 and W4 bands }
 We constructed the ensemble structure functions in four WISE bands as shown in  Figure 2. We fit the results  by four-parameter expression (3) in Wang\& Shi(2019) by the least square method, then obtain the slope $\beta=$ $2.75\pm0.19$ and $2.90\pm0.31$, respectively. The $\beta$ values suggest the ensemble SFs in W1 and W2 bands deviate from the damped random walk model ($\beta=1$). 
Koz{\l}owski et al.(2016) analyzed the mid-IR variability of 1500 AGNs based on the The Decadal IRAC Bo{\"o}tes Survey (DIBS) and its predecessor the Spitzer Deep, Wide-Field Survey (SDWFS) 
 by single power-law SF, which suggests the slope to be $\gamma \approx 0.45$ ($\beta\approx0.90$).  
It is obvious that our SF slopes are steeper than that in Koz{\l}owski et al.(2016). While each of their objects only have 5 epoch observations, nearly 2/5 of our objects have more than 10 epochs with on average denser cadence. As verified by Wang \& Shi (2019), the SF slope is sensitive to the sampling of a light curve.  As shown in the lower panel of Figure 2, the SFs in W3 and W4 bands do not show rising trends due to short time baselines.   

\begin{center}
		\begin{figure*}[ht]
			\includegraphics[width=62mm]{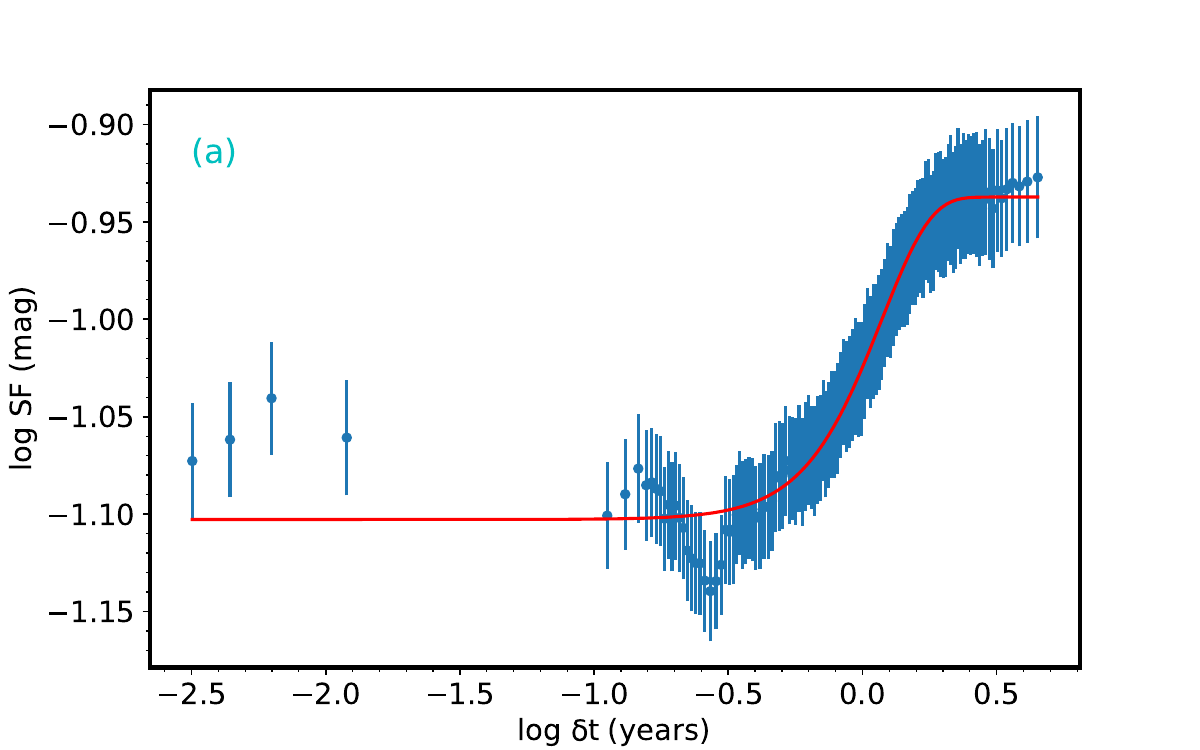}
			\includegraphics[width=62mm]{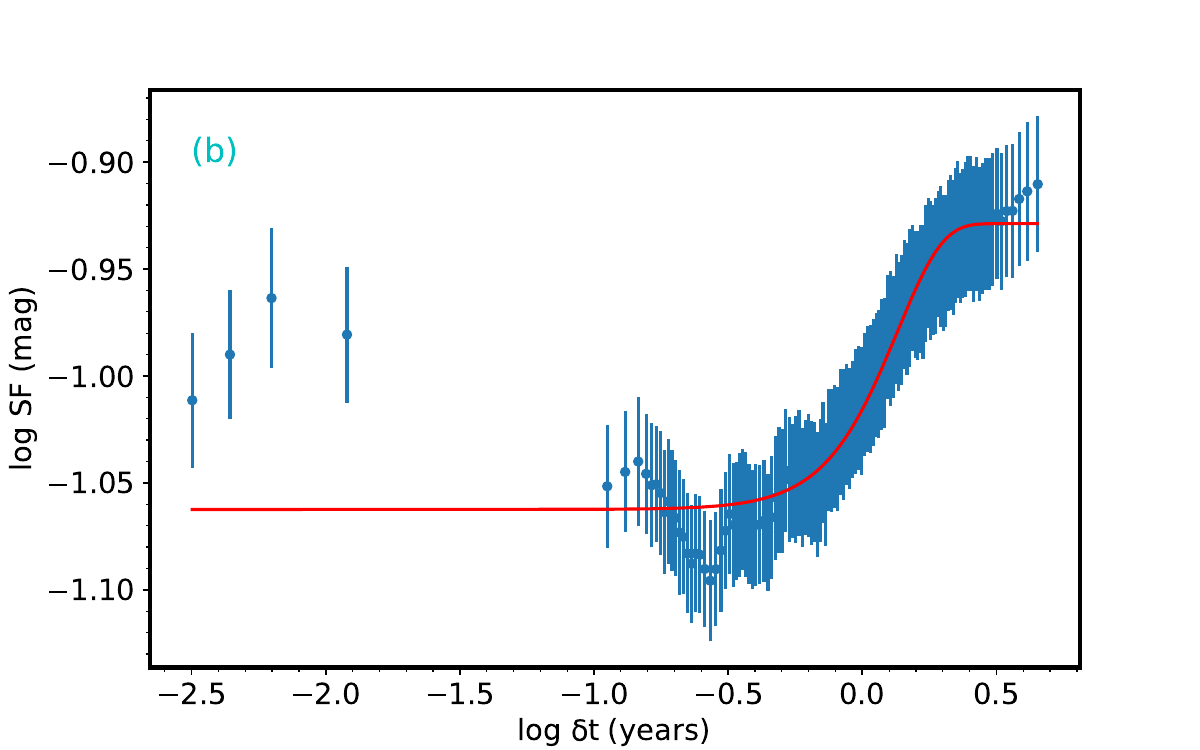}
		\end{figure*}
		\vspace{0.5mm}
		\begin{figure*}[ht] 
			\includegraphics[width=60mm]{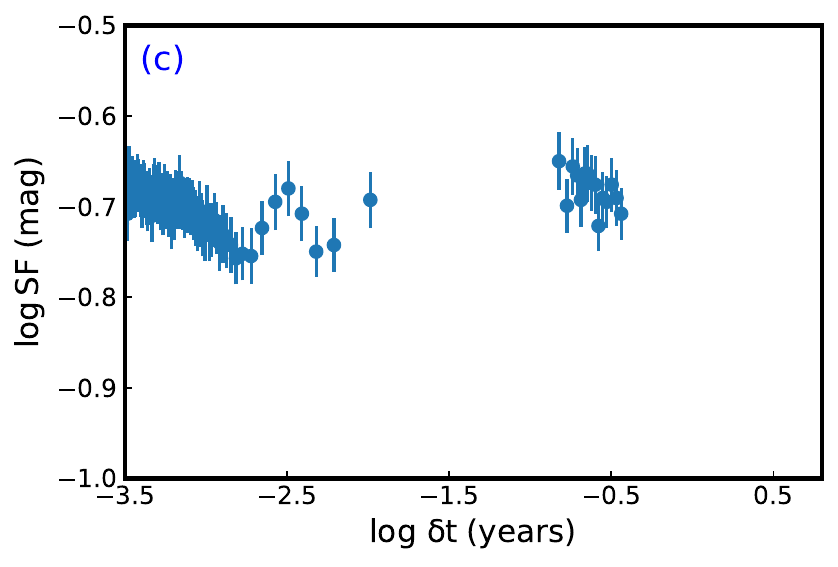}
			\includegraphics[width=60mm]{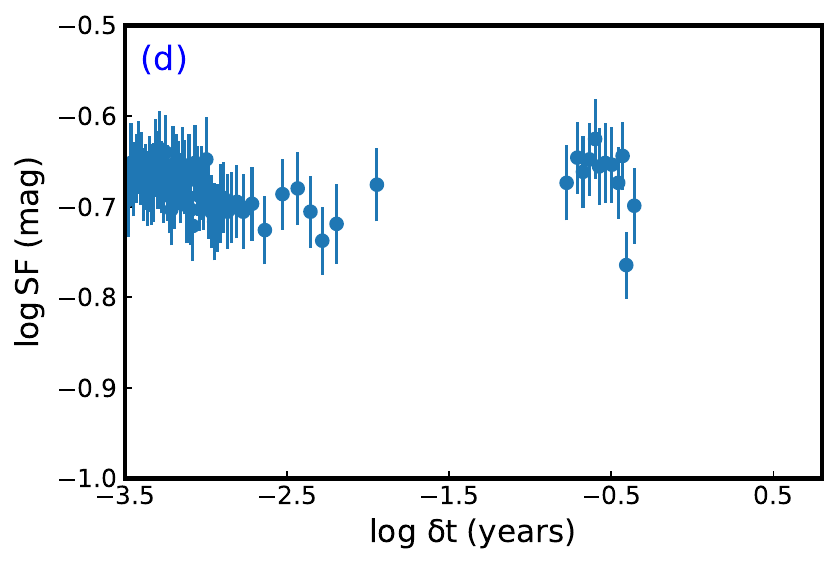}
		\caption{ The result of ensemble structure function in W1 (a), W2 (b), W3 (c) and W4 (d) bands }	
		\end{figure*}
\end{center}

\subsection{ The relation between the ensemble structure functions in W1 and W2 bands and redshift, bolometric luminosity, Eddington ratio as well as the black hole mass    } 
In Figure 3, we divide the sample into 10 bins in redshift (the bin size of redshift = 0.55) and 10 bins in bolometric luminosity (the bin size of luminosity = 0.45) to have in total 100 cells. We calculate the ensemble structure function in each cell and use the SF ($\delta t$ = 1 yr) as the variability amplitude. The timescale at one year is suitable because it is far away from the noise region in the short time lag and the observations are rich at this timescale. The number of quasars in each cell is presented in Figure 4. As shown in Figure 3, the SF ($\delta t$ = 1 yr) in the W1 band is about 0.1 mag, which is close the SF at $\delta t$ = 2 yr in Koz{\l}owski et al. (2010, 2016). No correlations are seen for this SF ($\delta t$ = 1 yr) with either redshift or the bolometric luminosity. The SF ($\delta t$ = 1 yr) in the W2 band show overall similar behavior to that in W1 band. 

We also analyze the relation between the SF ($\delta t$ = 1 yr) and the black hole mass as well as the  Eddington ratio. The distribution of the Eddington ratio and the black hole mass is shown in Figure 5.   
We also divide it into 10 bins in the black hole mass (the bin size = 0.3) from 7.3 to 10.3 and 10 bins
in the Eddington ratio (the bin size = 0.3) from -2.3 to 0.7. In each cell, the SF ($\delta t$ = 1 yr) in 
ensemble structure function is presented as the variability amplitude. As shown in Figure  6, no correlations are found for the SF ($\delta t$ = 1 yr) with either Eddington ratio or the black hole mass.

\begin{figure}[htbp]
\centering 
\resizebox{8cm}{5cm}
{\includegraphics[width=0.5\textwidth,trim= 0 30 0 70, clip]{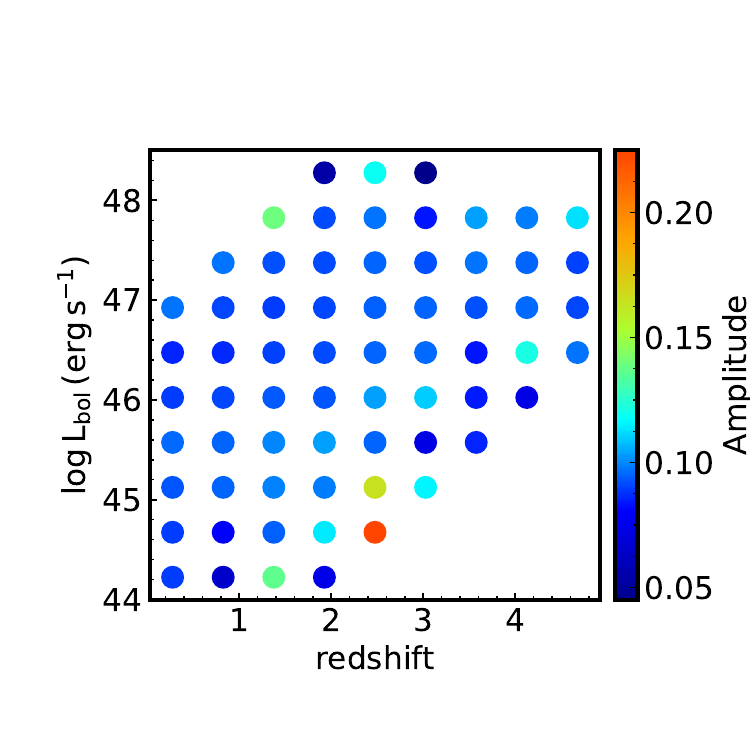}}
\resizebox{8cm}{5cm}
{\includegraphics[width=0.5\textwidth,trim= 0 30 0 70, clip]{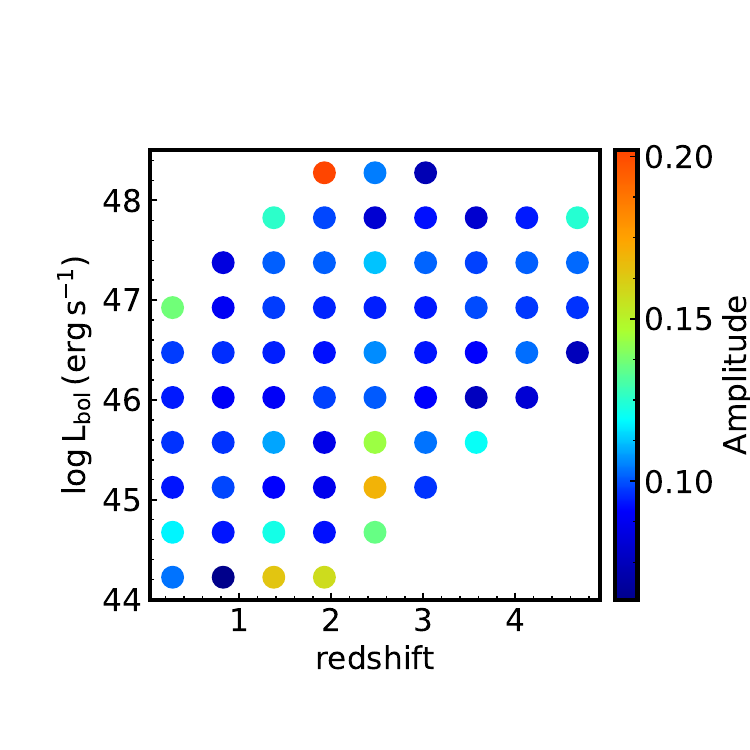}} 
\caption{ The analytical results in W1 band (upper panel) and W2 band (lower panel) by structure function.}\label{Fig:radio_var}. 
\end{figure} 

\begin{figure}[htbp]
\centering 
\resizebox{8cm}{6cm}{\includegraphics[width=0.5\textwidth,trim= 30 30 0 30, clip]{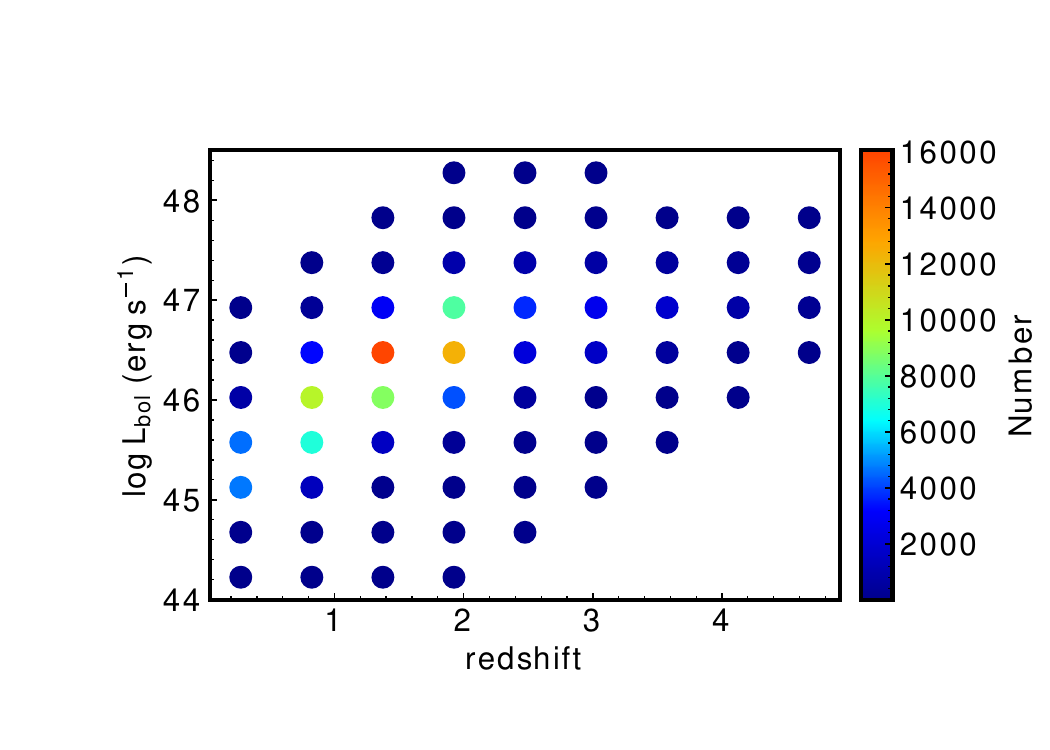}}
\caption{ The object number in W1 and W2 bands.}\label{Fig:radio_var}. 
\end{figure}

\begin{figure}[htbp]
\centering 
\resizebox{8cm}{6cm}{\includegraphics[width=0.5\textwidth,trim= 0 0 0 30, clip]{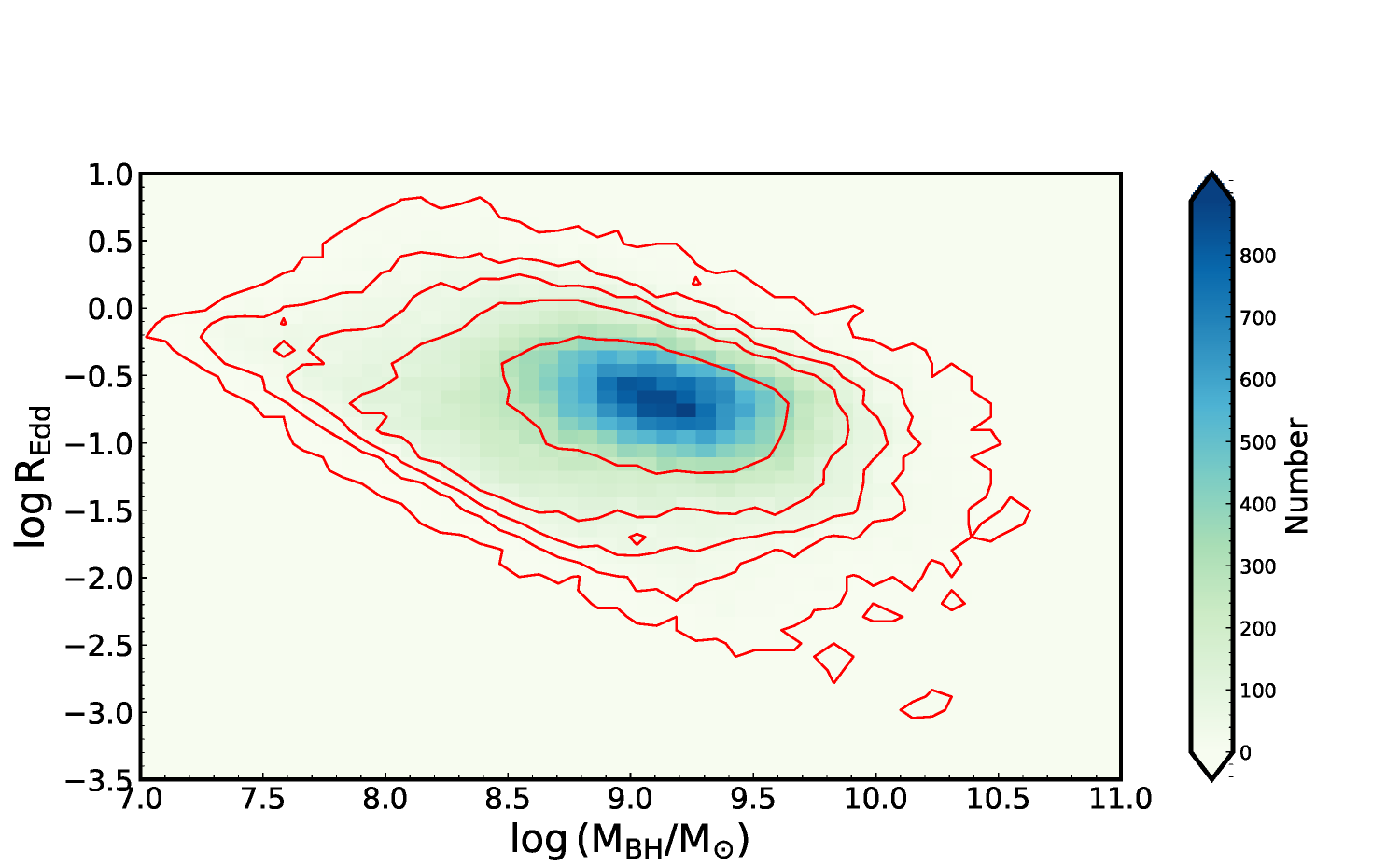}}
\caption{ The distribution of Eddington ratio and the black hole mass of the final quasar sample.}\label{Fig:radio_var}. 
\end{figure}

 \begin{figure}[htbp]
\centering 
\resizebox{8cm}{5cm}
{\includegraphics[width=0.5\textwidth,trim= 0 30 0 70, clip]{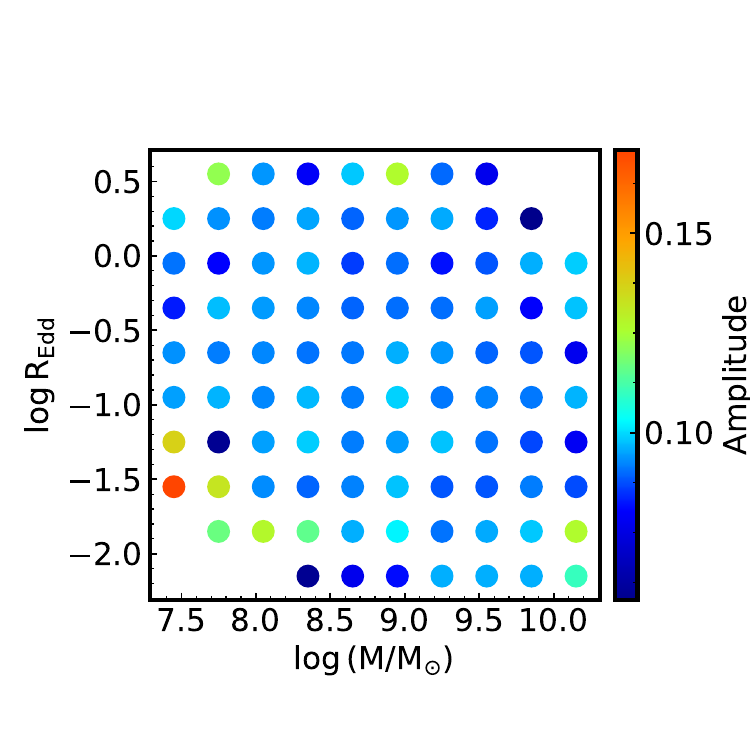}}
\resizebox{8cm}{5cm}
{\includegraphics[width=0.5\textwidth,trim= 0 30 0 70, clip]{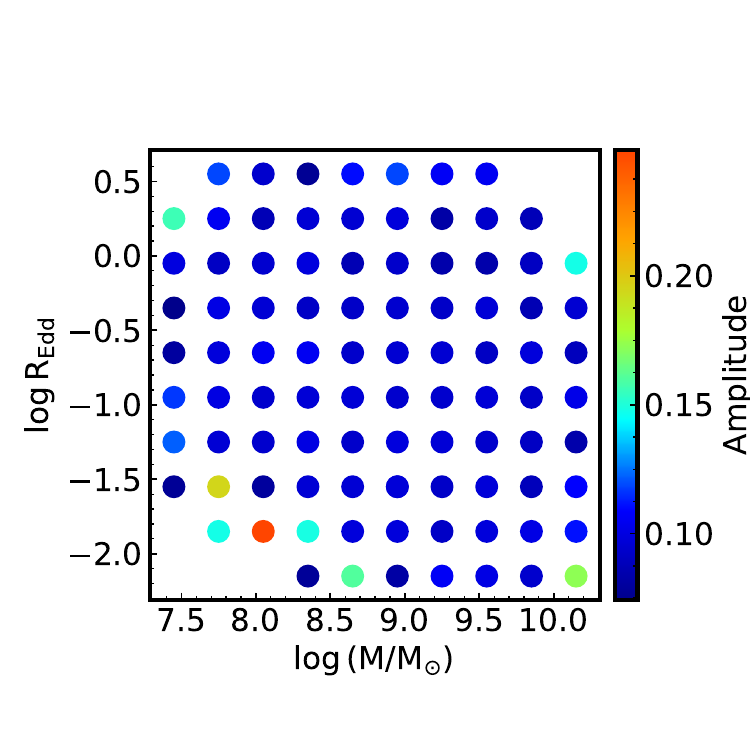}} 
\caption{ The relation between the variability amplitude (the SF ($\delta t$ = 1 yr)) in W1 band (upper panel) and W2 band (lower panel) and Edding ratio as well as black hole mass.}\label{Fig:radio_var}. 
\end{figure}

\subsection{ The relation between mid-infrared and optical variability }

\begin{figure}[t]
\centering 
\resizebox{8cm}{5cm}
{\includegraphics[width=0.5\textwidth,trim= 0 0 0 70, clip]{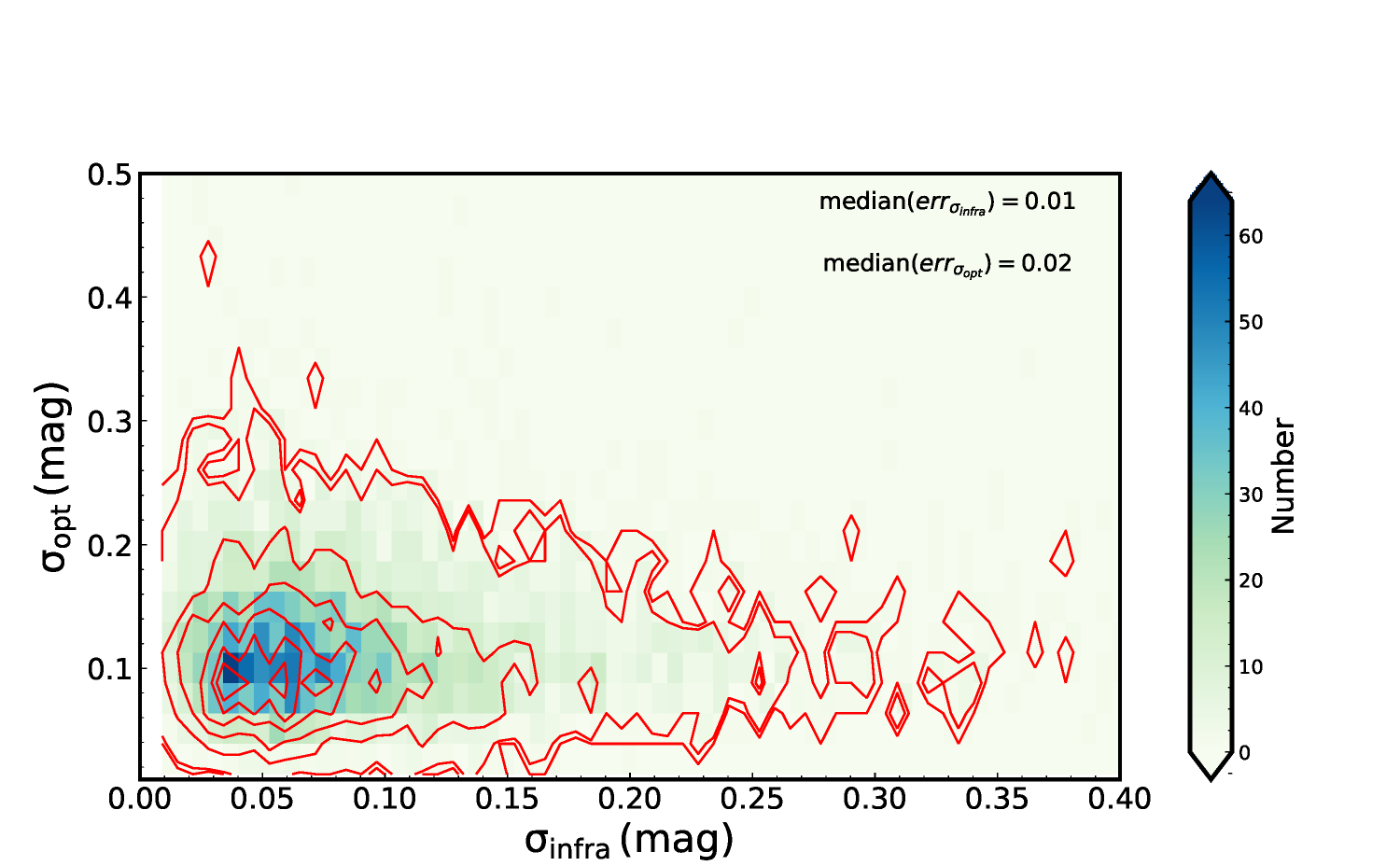}}
\resizebox{8cm}{5cm}
{\includegraphics[width=0.5\textwidth,trim= 0 0 0 70, clip]{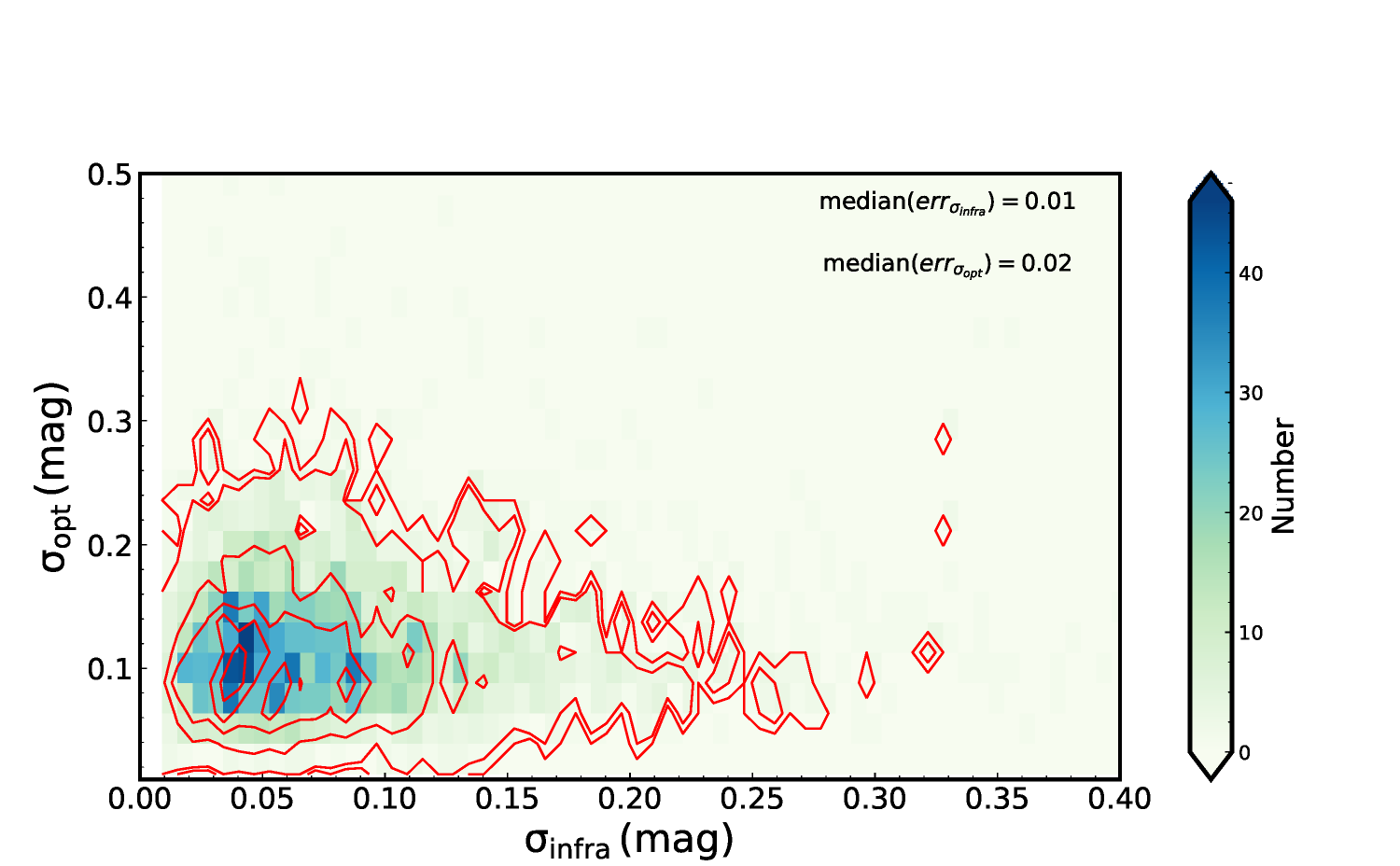}}
\caption{ The relation between mid-infrared and optical variability. The upper is the $\sigma_m$ in W1 band and the lower is the $\sigma_m$ in W2 band. }\label{Fig:radio_var}. 
\end{figure}

Based on the variability amplitude ($\sigma_m$) in Section 3.2, we further investigate the relation between mid-infrared and optical variability amplitude. We obtain the optical light curve from the SDSS Stripe 82 data sets from MacLeod et al. (2012), including $\sim$9000 quasars with $\sim$60 epoch observations and the time baseline of $\sim$ 9 years which is similar to that in mid-IR band.
 We choose r-band data sets to calculate the $\sigma_m $ due to its higher SNR, and more available data sets than that in u, g, i and z bands.   
At last, we obtain 3943 quasars with measurements of the variability amplitude in both W1 band and r band, 2856 quasars with both W2 band and r band. The correlations of the variability amplitudes between mid-IR and optical bands are presented in Figure 7. The variability amplitude in W1 band is in general less than 0.4 mag and shows no relationship with that in the r-band. Again, the W2 band shows similar behaviors. 
The most optical variable objects shows very small amplitudes of mid-IR variability in which the torus may 
largely evaporate due to the past high phase, thus the rest dust is far away from the black hole (Jiang et al.2017,2019). 
We assume the lag is simply determined by the torus inner radius ($\rm R_{sub} =0.5L_{46}^{0.5}(1800K/T_{sub})^{2.6} $pc), if the bolometric luminosity is $\rm 10^{46} erg s^{-1}$ and the temperature 1800 K, thus the time lag is about 2 yrs. The optical data sets from Stripe 82 coverage is from September 1998 to November 2007, and the mid-IR data sets is from January 2010 to April 2019.  Hence, the short time lag and different time coverage may weaken the correlation between optical band and mid-infrared band. 
 The lack of apparent relationships of the variability amplitude between W1/W2 and optical bands may also be caused by several factors such as the time lag between the optical and mid-IR emission, the smoothing effect in the mid-IR due to its extended size.

\subsection{ The color relation between mid-infrared band and optical band }
In order to further check the relation between mid-infrared band and optical band, we investigate the average color relation between W1-W2 and g-r. The color index g-r is from the SDSS Stripe 82 data sets. 
The result is presented in Figure 8. Most of the g-r is smaller than 0.5 mag. Most of the W1-W2 is between 
0.8 mag and 1.6 mag. The W1-W2 become flat trend with increasing g-r, which suggest no obvious correlation between W1-W2 and g-r in the DR7 quasar sample. Yang et al. (2018) analyzed the color variation of 21 changing-look AGNs, confirmed bluer-when-brighter trend in optical band and redder-when-brighter trend in mid-infrared band. No correlation between two bands may be also caused by the time lag between two bands and the smoothing effect in the mid-IR band.
 
\begin{figure}[htbp]
\centering 
\resizebox{8cm}{5cm}
{\includegraphics[width=0.5\textwidth,trim= 0 0 0 70, clip]{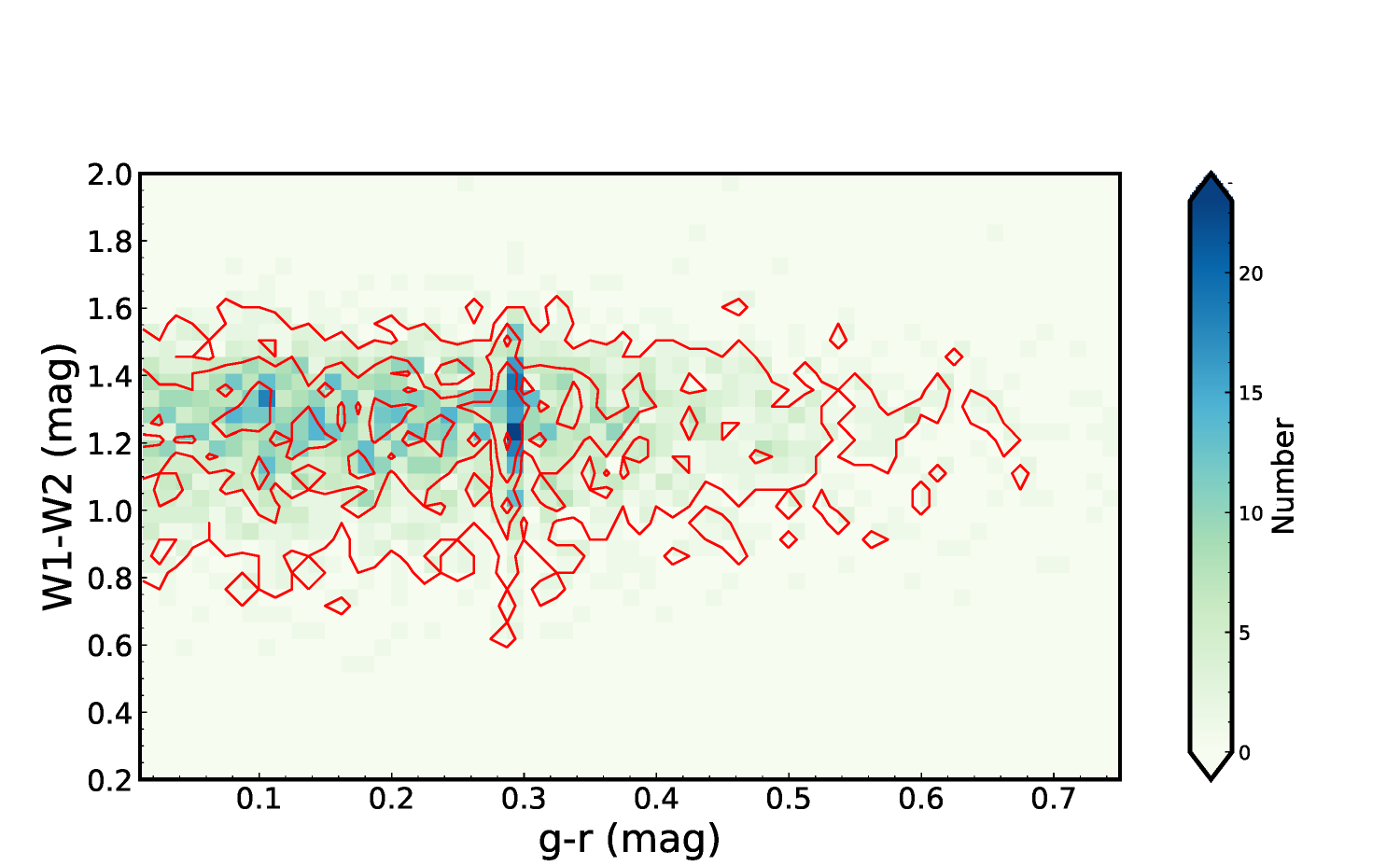}}
\caption{ The relationship between mid-infrared color (W1-W2) and optical color (g-r). }
\label{Fig:radio_var}. 
\end{figure}

\subsection{ The relation between the variability amplitude in mid-infrared and radio luminosity at 6 cm }
 In order to further examine whether there are some contributions from synchrotron emission in the mid-infrared band, we select a sub-sample of radio-loud quasars (the radio loudness above 10) from the Faint Images of the Radio Sky at Twenty-Centimeters (FIRST) radio survey (White et al. 1997). The radio loudness is defined as the ratio of the radio flux density at rest frame 6 cm (1.4 GHz) and the optical flux density at rest frame 2500 \AA ~(Shen et al. 2011). In order to further enhance the reliability, we exclude objects of $\sigma_m =0$ or the errors of $\sigma_m$ larger than 0.1 mag in the sub-sample. At last, we obtain 2949 objects that have  variability amplitude $\sigma_m$ in W1 band and 2008 objects in W2 band. The relationships between luminosity in 1.4 GHz and variability amplitude $\sigma_m$ in W1/W2 band are presented in Figure 9. No correlations are seen for both plots. We further select the objects between the redshift $0.0< z <0.5$, 660 and 548 objects in W1 band and W2 band, respectively. The results are shown in Figure 10 and also no  correlations are found.   
 Jiang et al. (2012) analyzed the mid-infrared variability of three radio-loud narrow-line Seyfert 1 galaxies using WISE data sets, and confirmed that they are blazars with jets close to our line of sight. Their infrared emission probably contains the synchrotron emission from the relativistic jet. Mao et al. (2018) showed $\gamma$-ray blazars had larger mid-IR variability than non-$\gamma$-ray blazars.  Bisogni et al. (2019) showed the infrared bump cannot be induced by synchrotron emission form the blazar-like objects, and found no difference between radio quiet and radio-loud objects in the stacked mid-infrared  spectral energy distribution. Rakshit et al. (2019) found a positive correlation between the variability amplitude and 1.4 GHz radio power by analyzing 63 radio-detected Narrow Line Seyfert 1 galaxies. From the SED view, the contribution from jet component in radio loud quasar is mainly in radio band (larger than $\sim$100 $\rm\mu m$), very few component in mid-IR band (Shang et al.2011). The synchrotron peak of SED in radio-loud NLS1 mainly distribute  from $10^{12}$ GHz (100 $\mu m$) to $10^{14}$ GHz (3 $\mu m$)(Abdo et al.(2009); Yao et al.(2015); Paliya et al.(2019)), then it's likely to be observed in mid-IR band.          
In contrast, our sub-sample is composed of radio-loud quasars, and find no obvious correlations between mid-infrared variability and radio luminosity, which are different results of Jiang et al. (2012) and Rakshit et al. (2010). This may be because of different galaxy types: narrow-line Seyfert in theirs vs. radio-loud quasars in ours. 

\begin{figure}[t]
\centering 
\resizebox{7.5cm}{5.5cm}
{\includegraphics{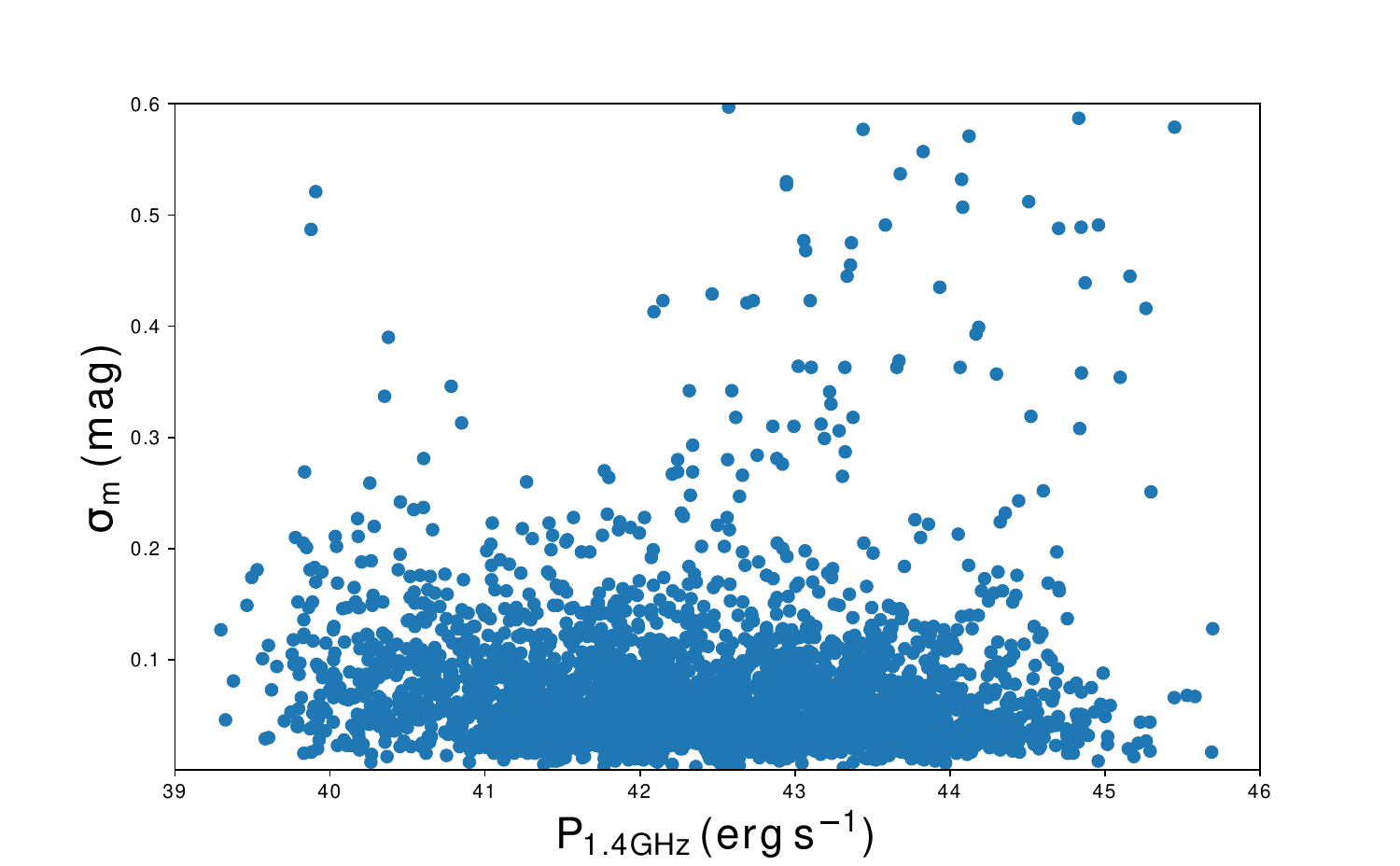}}
\resizebox{7.5cm}{5.5cm}
{\includegraphics{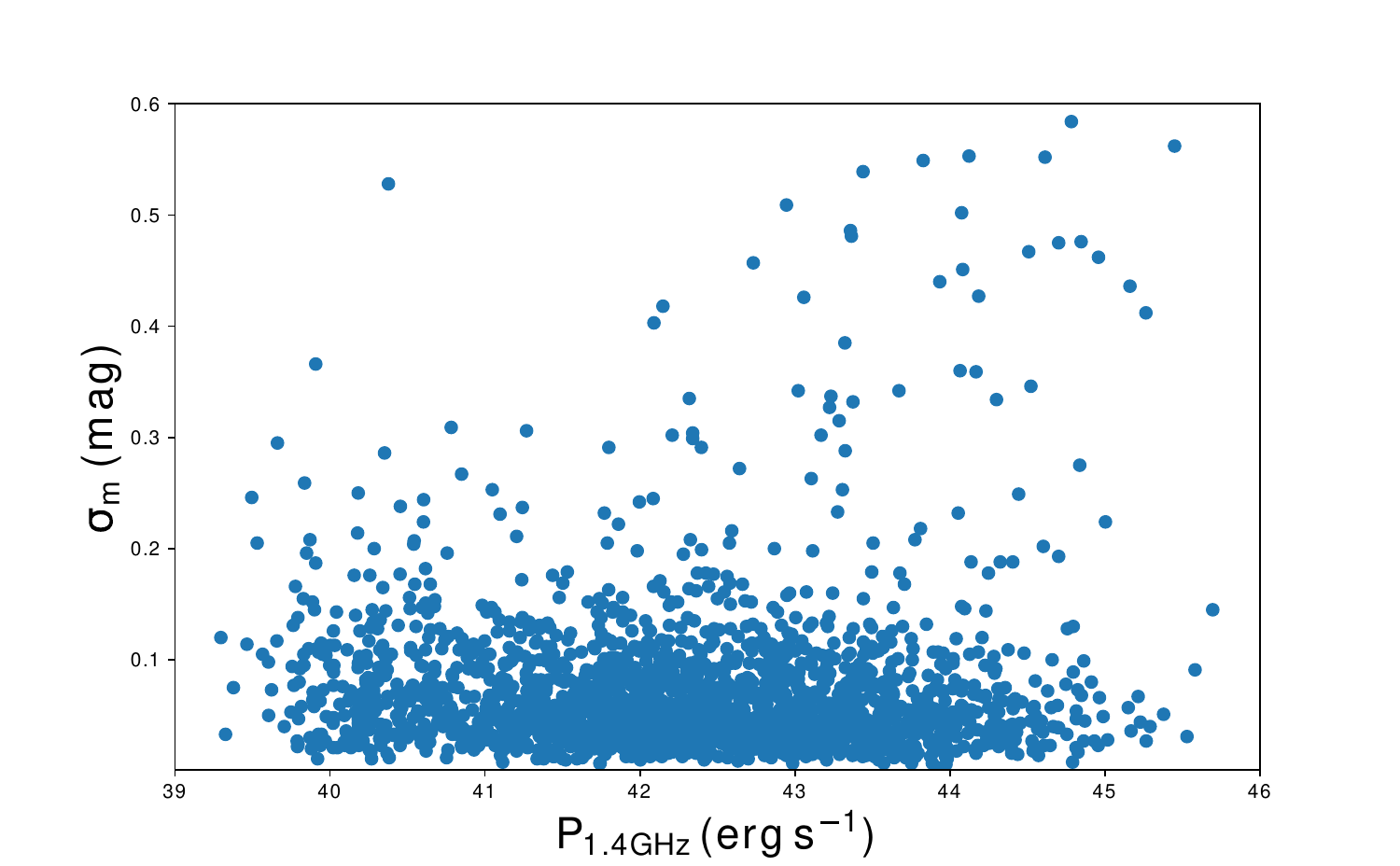}}
\caption{ The relation between the luminosity at 1.4 GHz and variation amplitude in mid-infrared W1
band (upper panel) and W2 band (lower panel).}\label{Fig:radio_var}. 
\end{figure}

\begin{figure}[t]
\centering 
\resizebox{7.5cm}{5.5cm}
{\includegraphics{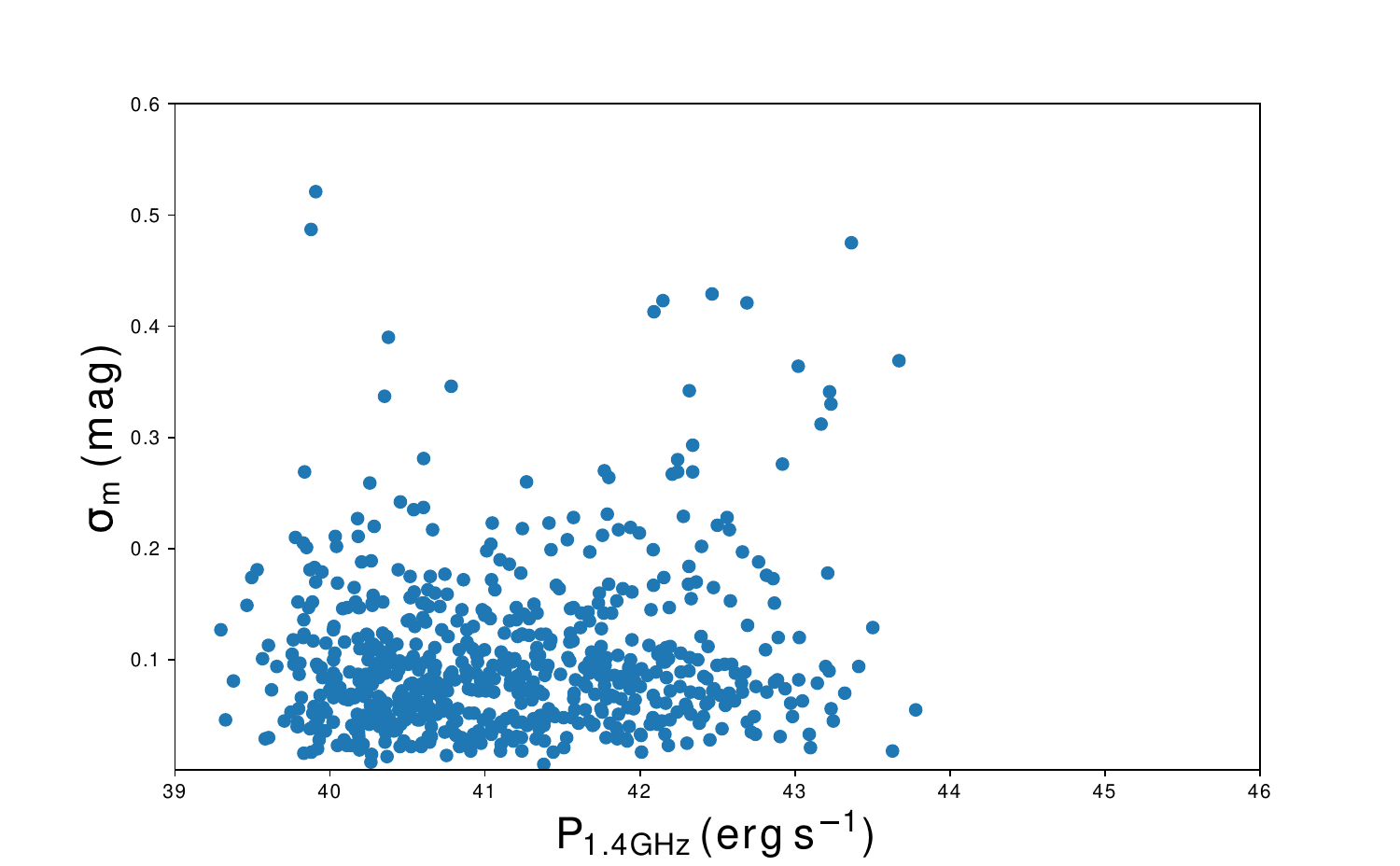}}
\resizebox{7.5cm}{5.5cm}
{\includegraphics{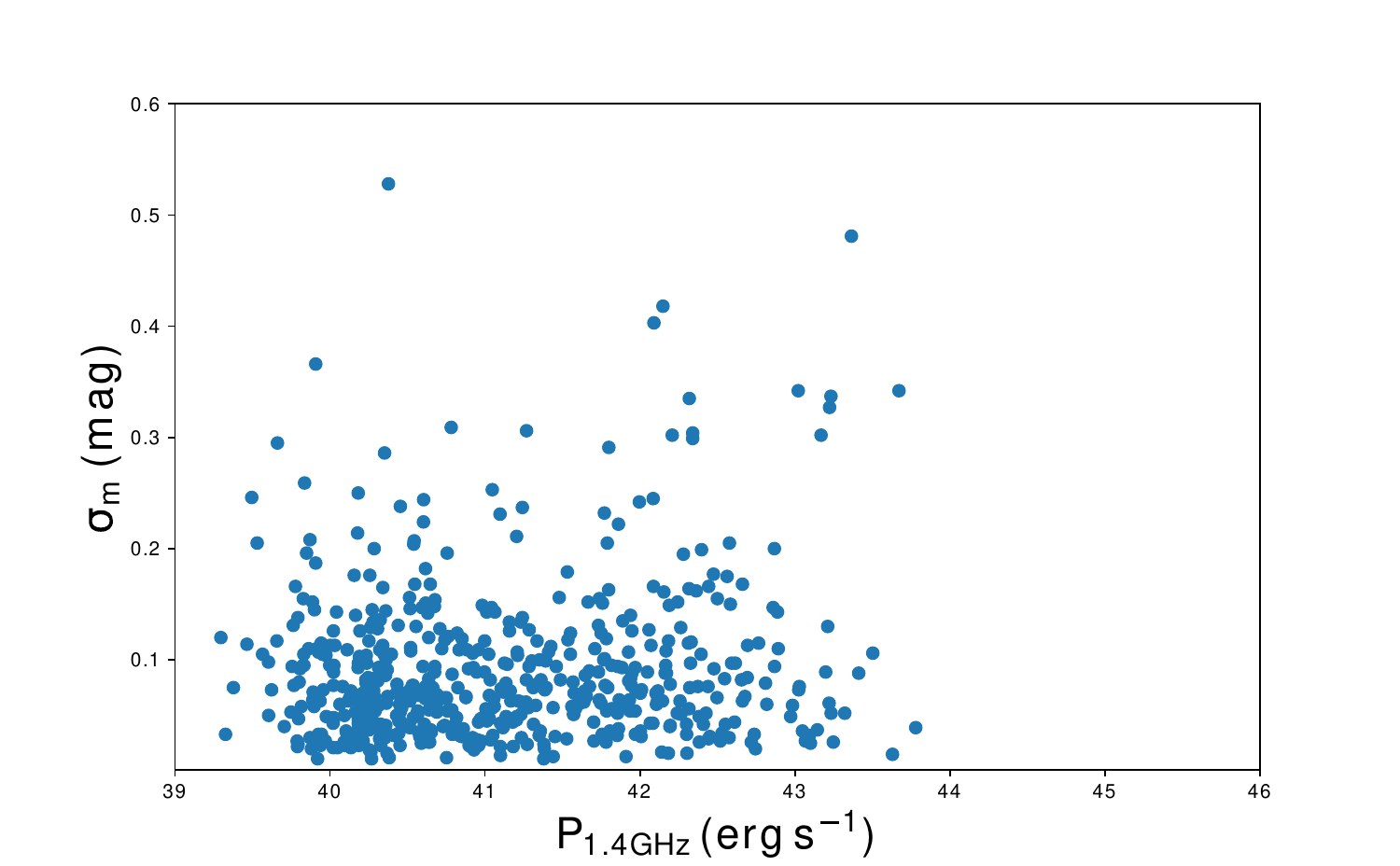}}
\caption{ The relation between the luminosity at 1.4 GHz and variation amplitude in mid-infrared W1
band (upper panel) and W2 band (lower panel) in the redshift $0.0 < z< 0.5$. }\label{Fig:radio_var}. 
\end{figure}

\section{Discussion} 

Koz{\l}owski et al. (2010) investigated the variability of objects with four epochs from the Spitzer Deep Wide-Field Survey  (SDWFS) in 3.6 and 4.5 $\mu$m bands by structure function method. They found the variability amplitude was $\sim$0.1 mag on the rest-frame timescale of two years and was anti-correlated with the luminosity. Koz{\l}owski et al. (2016) further added the fifth epoch from the follow-up the Decadal IRAC Bo{$\rm\ddot{o}$}tes Survey (DIBS) of the SDWFS, extending the time baseline to $\sim$ 10 years, and 
confirmed the above  anti-correlation. In this work, we find that the SF ($\delta t$ = 1 yr) is about 0.1 mag, which is consistent with the results in Koz{\l}owski et al. (2010, 2016). But the SF slopes in our result are steeper than that in Koz{\l}owski et al. (2010, 2016) which may be due to different cadence and number of observational epoch. We don't find obvious correlation between the bolometric luminosity and SF ($\delta t$ = 1 yr) in W1 and W2 bands. 

The Eddington ratio is usually considered to be the main driver of variability in quasars. Kelly (2009) analyzed about 100 radio quiet quasars and obtained the anti-correlation ($-0.25\pm0.22$) with the relation of log $\sigma^2$ and log L/LEdd. Rakshit\& Stalin (2017) used the data sets from Catalina Real Time Transient Survey to analyze the optical variabilty of 5080 Seyfert galaxies, and found the variability amplitude ($\sigma_m$) was anti-correlated with logL/LEdd, with the correlation coefficient $-0.16$ and p value $10^{-30}$. Rakshit et al.(2019) analyzed the mid-variability of 492 NLS1 galaxies using WISE data sets, and found the correlations of $\sigma_m$ in W1 and W2 bands were anti-correlated with log L/LEdd with the correlation coefficient (p value) are $-0.14(10^{-3})$ and $-0.21(10^{-6})$, respectively. The similar trends with Eddington ratio indicate the variability in mid-infrared band may be correlated with that in optical band.      
 
However, our results in Section 4.3 and 4.4 indicate that the correlations are not obvious. The variation in the mid-infrared is smoothed out by the extended dust distribution, hence the intrinsic relation of variation amplitude $\sigma_m$ between optical and infrared band may be smoothed out.  

Rakshit et al.(2019) found a positive correlation between the variability amplitude $\sigma_m$ in mid-IR W1/W2 band and radio luminosity from 63 radio-detected narrow-line Seyfert 1 galaxies. Their work inspires us to investigate the relation between mid-infrared variability and radio luminosity for radio-loud quasar sample.  
The results indicate no obvious correlations. It means the mid-infrared emission from jet contribution is weak in radio-loud quasars which is consistent with the view of Bisogni et al. (2019), and  implies that the emission of radio-loud quasars in mid-infrared band is dominated by the thermal emission from  dusty torus.

\section{Conclusions}
In this work, based on the multi-epoch observations from WISE and NEOWISE, we investigate the variability  characteristics of the SDSS DR7 quasars in W1,W2,W3 and W4 bands. The results are listed as follows.

(1). By means of SF method, we find that SF ($\delta t$ = 1 yr) shows no obvious correlation with the  bolometric luminosity, the black hole mass and the Eddington ratio. The SF slopes in W1 and W2 bands are steeper than that in Koz{\l}owski et al.(2016) which may be caused by different cadence and observational number. 

(2). We further investigate the relation of variability amplitude between mid-IR band and optical band, and no obvious correlation is found. The relation between W1-W2 and g-r is also investigated and shows no correlation . We think the mid-infrared emission of quasars may be smoothed out by the extended dust distribution, thus leading to no obvious correlation. 

(3). Based on the radio-loud quasar sub-sample, we analyze the relation between the variability amplitude in mid-infrared  band and the radio luminosity at 6 cm. No obvious correlations are found, which indicates that  the mid-infrared emission from synchrotron radiation of relativistic jet is weak.

\begin{acknowledgements}
 We thank the anonymous referees for helpful suggestions that improved the manuscript. We would like to thank Ge Xue from Nanjing university give us many useful help in drawing the figure by Python. We also thank Chen Yongyun and Yu Xiaoling from Nanjing university for their useful comments and discussion when preparing this paper. This publication makes use of data products from the Wide-field Infrared Survey Explorer, which is a joint project of the University of California, Los Angeles, and the Jet Propulsion Laboratory/California Institute of Technology, funded by the National Aeronautics and Space Administration, and also makes use of data products from the Near-Earth Object Wide-field Infrared Survey Explorer (NEOWISE), which is a project of the Jet Propulsion Laboratory/California Institute of Technology. NEOWISE is funded by the National Aeronautics and Space Administration. 
This work is supported by the National Key R$\&$D Program of China (No. 2017YFA0402704, No. 2018YFA0404502), the National Natural Science Foundation of China (NSFC grants 11733002 and 11773013), the Excellent Youth Foundation of the Jiangsu Scientific Committee (BK20150014). W.H.T acknowledges support from the Science and Technology supporting Program in Langfang city (No.2018011005). 
\end{acknowledgements}


\begin{thebibliography}{999}
\bibitem[Abdo et al.(2009)]{2009ApJ...707L.142A} Abdo, A.~A., Ackermann, M., Ajello, M., et al.\ 2009, \apjl, 707, L142
\bibitem[Ai et al.(2010)]{2010ApJ...716L..31A} Ai, Y.~L., Yuan, W., Zhou, H.~Y., et al.\ 2010, \apjl, 716, L31 
\bibitem[Ai et al.(2013)]{2013AJ....145...90A} Ai, Y.~L., Yuan, W., Zhou, H., et al.\ 2013, \aj, 145, 90 
\bibitem[Bauer et al.(2009)]{2009ApJ...696.1241B} Bauer, A., Baltay, C., Coppi, P., et al.\ 2009, \apj, 696, 1241
\bibitem[Bisogni et al.(2019)]{2019MNRAS.485.1405B} Bisogni, S., Lusso, E., Marconi, A., et al.\ 2019, \mnras, 485, 1405
\bibitem[Emmanoulopoulos et al.(2010)]{2010MNRAS.404..931E} Emmanoulopoulos, D., McHardy, I.~M., \& Uttley, P.\ 2010, \mnras, 404, 931 
\bibitem[Hughes et al.(1992)]{1992ApJ...396..469H} Hughes, P.~A., Aller, H.~D., \& Aller, M.~F.\ 1992, \apj, 396, 469
\bibitem[Jarrett et al.(2011)]{2011ApJ...735..112J} Jarrett, T.~H., Cohen, M., Masci, F., et al.\ 2011, \apj, 735, 112
\bibitem[Jiang et al.(2012)]{2012ApJ...759L..31J} Jiang, N., Zhou, H.-Y., Ho, L.~C., et al.\ 2012, \apjl, 759, L31 
\bibitem[Jiang et al.(2017)]{2017ApJ...850...63J} Jiang, N., Wang, T., Yan, L., et al.\ 2017, \apj, 850, 63
\bibitem[Jiang et al.(2019)]{2019ApJ...871...15J} Jiang, N., Wang, T., Mou, G., et al.\ 2019, \apj, 871, 15
\bibitem[Kelly et al.(2009)]{2009ApJ...698..895K} Kelly, B.~C., Bechtold, J., \& Siemiginowska, A.\ 2009, \apj, 698, 895
 \bibitem[Koshida et al.(2014)]{2014ApJ...788..159K} Koshida, S., Minezaki, T., Yoshii, Y., et al.\ 2014, \apj, 788, 159 
 \bibitem[Koz{\l}owski et al.(2010)]{2010ApJ...716..530K} Koz{\l}owski, S., Kochanek, C.~S., Stern, D., et al.\ 2010, \apj, 716, 530
 \bibitem[Koz{\l}owski et al.(2016)]{2016ApJ...817..119K} Koz{\l}owski, S., Kochanek, C.~S., Ashby, M.~L.~N., et al.\ 2016, \apj, 817, 119
 \bibitem[MacLeod et al.(2012)]{2012ApJ...753..106M} MacLeod, C.~L., Ivezi{\'c}, {\v{Z}}., Sesar, B., et al.\ 2012, \apj, 753, 106
\bibitem[Mandal et al.(2018)]{2018MNRAS.475.5330M} Mandal, A.~K., Rakshit, S., Kurian, K.~S., et al.\ 2018, \mnras, 475, 5330 
\bibitem[Matthews \& Sandage(1963)]{1963ApJ...138...30M} Matthews, T.~A., \& Sandage, A.~R.\ 1963, \apj, 138, 30 
\bibitem[Mao et al.(2018)]{2018Ap&SS.363..167M} Mao, L., Zhang, X., \& Yi, T.\ 2018, \apss, 363, 167
\bibitem[McHardy et al.(2006)]{2006Natur.444..730M} McHardy, I.~M., Koerding, E., Knigge, C., Uttley, P., \& Fender, R.~P.\ 2006, \nat, 444, 730 
\bibitem[Nenkova et al.(2008)]{2008ApJ...685..160N} Nenkova, M., Sirocky, M.~M., Nikutta, R., et al.\ 2008, \apj, 685, 160
\bibitem[Padovani et al.(2017)]{2017A&ARv..25....2P} Padovani, P., Alexander, D.~M., Assef, R.~J., et al.\ 2017, \aapr, 25, 2 
\bibitem[Paliya et al.(2019)]{2019ApJ...872..169P} Paliya, V.~S., Parker, M.~L., Jiang, J., et al.\ 2019, \apj, 872, 169
\bibitem[Pancoast et al.(2011)]{2011ApJ...730..139P} Pancoast, A., Brewer, B.~J., \& Treu, T.\ 2011, \apj, 730, 139 
\bibitem[Peterson(2004)]{2004IAUS..222...15P} Peterson, B.~M.\ 2004, The Interplay Among Black Holes, Stars and ISM in Galactic Nuclei, 15
\bibitem[Rakshit, \& Stalin(2017)]{2017ApJ...842...96R} Rakshit, S., \& Stalin, C.~S.\ 2017, \apj, 842, 96
\bibitem[Rakshit et al.(2019)]{2019MNRAS.483.2362R} Rakshit, S., Johnson, A., Stalin, C.~S., et al.\ 2019, \mnras, 483, 2362
\bibitem[Rees(1984)]{1984ARA&A..22..471R} Rees, M.~J.\ 1984, \araa, 22, 471 
\bibitem[Sesar et al.(2007)]{2007AJ....134.2236S} Sesar, B., Ivezi{\'c}, {\v{Z}}., Lupton, R.~H., et al.\ 2007, \aj, 134, 2236
\bibitem[Shakura \& Sunyaev(1973)]{1973A&A....24..337S} Shakura, N.~I., \& Sunyaev, R.~A.\ 1973, \aap, 24, 337 
\bibitem[Shang et al.(2011)]{2011ApJS..196....2S} Shang, Z., Brotherton, M.~S., Wills, B.~J., et al.\ 2011, \apjs, 196, 2
\bibitem[Shen et al.(2011)]{2011ApJS..194...45S} Shen, Y., Richards, G.~T., Strauss, M.~A., et al.\ 2011, \apjs, 194, 45
\bibitem[Simonetti et al.(1985)]{1985ApJ...296...46S} Simonetti, J.~H., Cordes, J.~M., \& Heeschen, D.~S.\ 1985, \apj, 296, 46
\bibitem[Sumi et al.(2005)]{2005MNRAS.356..331S} Sumi, T., Wo{\'z}niak, P.~R., Eyer, L., et al.\ 2005, \mnras, 356, 331
\bibitem[Sun et al.(2015)]{2015ApJ...811...42S} Sun, M., Trump, J.~R., Shen, Y., et al.\ 2015, \apj, 811, 42
\bibitem[Ulrich et al.(1997)]{1997ARA&A..35..445U} Ulrich, M.-H., Maraschi, L., \& Urry, C.~M.\ 1997, \araa, 35, 445 
\bibitem[Urry \& Padovani(1995)]{1995PASP..107..803U} Urry, C.~M., \& Padovani, P.\ 1995, \pasp, 107, 803 
\bibitem[Vanden Berk et al.(2004)]{2004ApJ...601..692V} Vanden Berk, D.~E., Wilhite, B.~C., Kron, R.~G., et al.\ 2004, \apj, 601, 692
\bibitem[Vestergaard, \& Peterson(2006)]{2006ApJ...641..689V} Vestergaard, M., \& Peterson, B.~M.\ 2006, \apj, 641, 689
\bibitem[Wagner \& Witzel(1995)]{1995ARA&A..33..163W} Wagner, S.~J., \& Witzel, A.\ 1995, \araa, 33, 163 
\bibitem[Wang, \& Shi(2019)]{2019Ap&SS.364...27W} Wang, H., \& Shi, Y.\ 2019, \apss, 364, 27
\bibitem[White et al.(1997)]{1997AAS...19110305W} White, R.~L., Becker, R.~H., Gregg, M.~D., et al.\ 1997, American Astronomical Society Meeting Abstracts 191, 103.05
\bibitem[Wright et al.(2010)]{2010AJ....140.1868W} Wright, E.~L., Eisenhardt, P.~R.~M., Mainzer, A.~K., et al.\ 2010, \aj, 140, 1868
\bibitem[Yang et al.(2018)]{2018ApJ...862..109Y} Yang, Q., Wu, X.-B., Fan, X., et al.\ 2018, \apj, 862, 109
\bibitem[Yao et al.(2015)]{2015MNRAS.454L..16Y} Yao, S., Yuan, W., Zhou, H., et al.\ 2015, \mnras, 454, L16



 
\end{thebibliography}
\end{document}